\documentclass[10pt, a4paper, onecolumn]{article}

\usepackage[numbers]{natbib}

\usepackage[T1]{fontenc}
\usepackage{lmodern}

\usepackage{natbib}
\usepackage[utf8]{inputenc} 
\usepackage{booktabs}       
\usepackage{amsfonts}       
\usepackage{nicefrac}       
\usepackage{microtype}      
\usepackage{xcolor}         
\usepackage{graphicx}
\usepackage{amsmath,amssymb,amsfonts}
\usepackage{bm}
\usepackage{url}

\usepackage{array,multirow,graphicx}

\usepackage{geometry}
\newgeometry{vmargin={15mm}, hmargin={12mm,17mm}} 

\title{One step closer to EEG based eye tracking}

\author{
	Wolfgang Fuhl
	\and
	Susanne Zabel
	\and
	Theresa Harbig
	\and
	Julia Astrid Moldt
	\and
	Teresa Festl Wiete
	\and
	Anne Herrmann Werner
	\and
	Kay Nieselt
	}

\date{}

\begin{document}
	
	\maketitle
	
	\begin{abstract}
		In this paper, we present two approaches and algorithms that adapt areas of interest We present a new deep neural network (DNN) that can be used to directly determine gaze position using EEG data. EEG-based eye tracking is a new and difficult research topic in the field of eye tracking, but it provides an alternative to image-based eye tracking with an input data set comparable to conventional image processing. The presented DNN exploits spatial dependencies of the EEG signal and uses convolutions similar to spatial filtering, which is used for preprocessing EEG signals. By this, we improve the direct gaze determination from the EEG signal compared to the state of the art by 3.5 cm MAE (Mean absolute error), but unfortunately still do not achieve a directly applicable system, since the inaccuracy is still significantly higher compared to image-based eye trackers. \\
		Link: \url{https://es-cloud.cs.uni-tuebingen.de/d/8e2ab8c3fdd444e1a135/?p=%2FEEGGaze&mode=list}
	\end{abstract}

	\section{Introduction}
	Eye tracking is a rising research field which finds more and more application areas like foveated rendering~\cite{walton2021beyond}, supportive diagnostics~\cite{ding2019classifying}, human computer interaction in games~\cite{sundstedt2012gazing}, driver observation in autonomous cars~\cite{vetturi2020use}, interaction technology for people with limitations~\cite{wanluk2016smart,dahmani2020intelligent}, automatic focus system for surgical microscopes~\cite{eivazi2018eyemic,fuhl2017fast}, marketing evaluation of websites or online shops~\cite{kamangar2020literature,munoz2019measuring}, and many more. The dominant approaches in eye tracking are image based due to multiple reasons. The main reason for image based eye tracking is that it is less invasive compared to other methods like the scleral coil~\cite{whitmire2016eyecontact} or EEG with blink frequencies~\cite{vettori2020combined}. In addition, image based eye tracking is very accurate but still has its limitations under challenging settings like near infrared reflections on glassesor in adequate lightning conditions~\cite{fuhl2015excuse}. In the past years there were multiple open source tools published~\cite{fuhl2022pistol,santini2017eyerectoo,UMUAI2020FUHL,kassner2014pupil,mantiuk2012yourself} which can be used freely for eye tracking. Additionally, multiple algorithms for pupil~\cite{ETRA2018FuhlW,ICCVW2019FuhlW,CAIP2019FuhlW}, eyelid~\cite{fuhl2022pistol}, and iris~\cite{WTDTWE092016,WTDTE022017,WTE032017,fuhl2022pistol} extraction were published which also lead to improved eye trackers from the industry~\cite{kassner2014pupil,tonsen2020high,gibaldi2017evaluation}. The disadvantages of image based eye tracking are that the worn (head mounted) eye tracker is invasive since you either have to wear glasses with integrated LEDs and cameras or the eye cameras are in your field of vision~\cite{valtakari2021eye}. This is due to the fact that high off-axial angles lead to a decrease in accuracy~\cite{swirski2012robust}. For remote eye tracking, the cameras can be placed everywhere in the room, but the distance to the subjects also decreases the accuracy~\cite{hosp2020remoteeye}. In addition, the whole setup has to be calibrated, and it is not guaranteed that no subject is concealed during the recording~\cite{fuhl2022groupgazer}.
	
	Today's research also starts to focus on EEG based eye tracking~\cite{nagel2019world,kastrati2021eegeyenet}. So far there are multiple approaches published which try to use EEG as visual interaction signal. The first approach uses high blink frequencies on graphical user interface elements like buttons. These frequencies are not perceived by the subject but can be extracted from the EEG signal since the visual input of a human is processed in the brain and, therefore, in the EEG data~\cite{nagel2019world}. Another approach is able to distinguish control fixations from visual exploration fixations. Therefore, it can be used to start actions when viewing a control element~\cite{shishkin2016eeg}. An approach to estimate the saccade direction with deep learning models was proposed in ~\cite{kastrati2021using} and improved to direct gaze estimation based on the EEG signal~\cite{kastrati2021eegeyenet}. While those approaches are still evaluated under laboratory conditions, an EEG based eye tracker would further reduce the invasiveness compared to a head mounted eye tracker. This is especially true for dry EEG setups with wireless communications~\cite{lopez2014dry}. 
	Our approach also focuses on the direct gaze estimation. For this, we propose a novel architecture which makes use of the spatial filtering usually applied to the EEG signal before it is processed. In our case, this spatial filtering is integrated into the model itself and learned during training. For a comparison to the state of the art we used the same training data as in ~\cite{kastrati2021eegeyenet}. Our model improves the gaze estimation accuracy by 3.5 cm as MAE (Mean absolute error) and has fewer parameters and a higher throughput compared to EEGEyeNet~\cite{kastrati2021eegeyenet}.

	\section{Related Work}
	For eye tracking and its core discipline, the gaze estimation, there are already multiple ways published in the literature. The world's first approach was the observation of participants, which was done by ophthalmologist Louis Émile Javal. He noticed that the does not move fluently while reading, but makes quick movements. 1908 Edmund Huey build the first eye tracker by adding a pointing device to a contact lens. While this was very intrusive, it opened up the way to more complex setups and systems for eye tracking. An alternative to the Edmund Huey eye tracker is the scleral coil which is a contact lens with an integrated circuit~\cite{whitmire2016eyecontact,massin2020development} or an integrated small magnet~\cite{bellizzi2022innovative}. The user itself is in an elector magnetic field and his eye movements, also move the contact lens. Based on this setup, a very accurate gaze estimation is possible. The biggest disadvantage of this approach is that the user has to wear a contact lens which is uncomfortable, especially for subjects, which are not used to contact lenses.
	
	An alternative to the electromagnetic field based eye tracking is the nowadays most popular way, which is the image based eye tracking. In this field, multiple approaches and solutions are published. The most basic approach is the iris and/or pupil tracking combined with a calibration, which maps the pupil/iris center to the target area~\cite{ramanauskas2006calibration,villanueva2004eye,duchowski2017eye}. Since the basic approach is very prone to shifts of the worn eye tracker, multiple improvements have been proposed. The most common shift robust approach is to project some near infrared glints onto the sclera~\cite{duchowski2017eye,blignaut2014mapping,hua2006video}. Afterwards, the glints as well as the pupil is detected and a vector between the centers is computed. This vector is less prone to eye tracker shifts but limits the field of view which can be tracked since the glints are not always visible~\cite{duchowski2017eye,blignaut2014mapping,hua2006video}. This approach also fails if there are additional sources of near infrared light, like direct sunlight. Here it is not possible anymore to detect the correct glints. An alternative to the glint based approach is a 3D eyeball model~\cite{swirski2013fully}. With the computed 3D eyeball, the optical vector can be used for the calibration, which is independent of eye tracker shifts. It is still prone to errors due to the computation of the 3D eyeball and has problems, especially outside the calibration area. The last and newest approach is the direct gaze estimation from the face~\cite{xu2018gaze} or eye images~\cite{zhang2015appearance,wood2016learning}. Here, modern deep neural networks are used and trained to estimate the gaze direction. While those approaches are still very inaccurate and struggle if the image domain is changed (Different cameras or NIR to RGB), they are still a fast and simple way to apply gaze estimation. They perform especially well if the resolution of the eye is very low, since feature extraction (Pupil, Iris, etc.) fails here.
	
	EEG (Electroencephalography) based eye tracking is a novel approach and uses brain signals to estimate the gaze location or perceived elements from a GUI (Graphical user interface) for example. One approach from this domain uses high blink frequencies on graphical user interface elements like buttons~\cite{nagel2019world}. These blink frequencies are not perceived by the subject, but they are measurable in the EEG signal~\cite{nagel2019world}. Each GUI element has its own frequency and based on this, they can be distinguished. Another approach for GUIs is to distinguish between visual exploration fixations and control or action fixations based on the EEG signal~\cite{shishkin2016eeg}. Based on this decision, buttons can be pressed, and the user is able to interact with a computer. In \cite{kastrati2021using} a deep learning model was proposed to estimate the saccade direction of the subject. This approach was further improved to real gaze estimation with a deep learning model, as well as other approaches like saccade direction and amplitude estimation were evaluated~\cite{kastrati2021eegeyenet}. While this is a promising way, all EEG research so far is conducted under laboratory conditions and using wet EEG electrodes, which consume a lot of time to place. With dry EEG devices~\cite{lopez2014dry}, the EEG based gaze estimation could find its way into the real world.
	
	\section{Method}
	Since the main work horse of our model is the spatial filtering convolution which we have created, we start with a small introduction to EEG based spatial filtering and afterwards explain how we came up with our design of the spatial filtering in a deep neural network. Spatial filtering of EEG data has a long history and there exist multiple basic approaches for it like common average reference, Laplacian filters, and principal component analysis~\cite{wu2017spatial}. More advanced approaches like the Independent Component Analysis, xDAWN algorithm, Common Spatial Patterns, and Canonical Correlation Analysis are the ones used nowadays for different application areas. The Independent Component Analysis decomposes the EEG signals into independent non-Gaussian signals for further processing. Similar to the Independent Component Analysis is the xDAWN algorithm, which additionally tries to increase the signal to signal plus noise ratio. For the Common Spatial Patterns approach, which is usually used in binary classification tasks, patterns between EEG signals of the two classes are searched with maximal difference in variance. The last approach is the Canonical Correlation Analysis, from which our idea steams from. Here, a linear transformation is searched to achieve the maximal correlation between EEG recordings. A linear transformation is a factor and an additive component, which is equal to a $1 \times 1$ convolution. This means our $1 \times 1$ convolution provides one learnable parameter per input signal of the EEG signals, and additionally one bias term per convolution. Therefore, we have a learnable spatial EEG filter.
	
	In our case, we have as input 129 EEG signals and from each signal 500 samples~\cite{kastrati2021eegeyenet} (With samples we refer to the data we receive from the EEG per timestamp). For each signal we learn one weight per convolution and one bias term for all signals per convolution.
	
	\begin{equation}
		\label{eq:spatialconv}
		Output_s=(\sum_{i=1}^{129} EEG_i * W_i ) + Bias
	\end{equation}
	
	Equation~\ref{eq:spatialconv} shows the computation of the convolution for one sample or timestamp $s$. $EEG_i$ is the raw EEG data of the electrode $i$ and $W_i$ is the weight of the convolution assigned to this electrode. $Bias$ is the learnable additive variable. Since the Canonical Correlation Analysis computes multiple of those linear transformations, we used 16 such convolutions, as can be seen in Table~\ref{tbl:archie}. Additionally, to the convolution which represents our linear transformation, we use batch normalization after the convolution. This is a normal concatenation of Layers in neural networks together with the following ReLu, but this also represents a linear transformation which is learned. Here, a moving average and the moving variance are computed online and applied together with two learnable parameters for scaling and shifting. This linear transformation can be seen as a global linear transformation between the 16 linear transformations learned via the convolutions. In addition, it reduces the training time of the neural network as well as it transforms the data optimally for the linearity breaking ReLu, which follows the batch normalization. But the last mentioned benefits of the batch norm are common and well known in neural networks.
	
	\begin{table}
		\centering
		\caption{Detailed description of our residual blocks. Normal residual blocks have either equally sized convolutions (Layer 2 and 3) or an ingoing as well as an outgoing $1 \times 1$ convolution. The parameter $N$ specifies the number of convolutions which are learned. In addition, the step size of the convolutions is 1 and the down scaling operation is only performed by the average pooling.}
		\label{tbl:residualblock}
		\begin{tabular}{c|l}
			Layer  & Type \\
			1  & Input size $ depth \times y-axis \times x-axis$ \\
			2  & $N$ Convolutions with size $depth \times 9 \times 1$, Batchnorm, ReLu \\
			3  & $N$ Convolutions with size $N \times 1 \times 1$, Batchnorm \\
			4  & Add input \\
			5  & ReLu \\
			6  & Average pooling with size $1 \times 2 \times 1$ \\
		\end{tabular}
	\end{table}
	Table~\ref{tbl:residualblock} shows the architecture of our residual blocks, from which we used two in our model. The first convolution, has a spatial dimension of $9 \times 1$ and the second convolution $1 \times 1$. This is uncommon in residual blocks since they usually have either equally sized convolutions or represent bottleneck blocks in which a $1 \times 1$ convolution is placed at the first and last position as well as a $3 \times 3$ convolution with a lower depth in the central position. For us the unequal convolution blocks worked best as can be seen in Table~\ref{tbl:design}. The idea itself steams from modern computer vision models which use convolutions~\cite{liu2022convnet} as well as from the transformers~\cite{dosovitskiy2020image} which separate the input into multiple larger segments and process them with fully connected layers which also represents larger convolutions. In our evaluations for the design decisions (Table~\ref{tbl:design}) we mention equally sized convolutions. This means that layer 2 and 3 use $9 \times 1$ convolutions.
	
	\begin{table}
		\centering
		\caption{Detailed description of our final model with the spatial filtering convolution (Layer 2) and the residual blocks with unequal convolutions. After the residual blocks, we use two fully connected layers, which is common for regression problems.}
		\label{tbl:archie}
		\begin{tabular}{c|l}
			Layer  & Type \\
			1  & Input size $ 129 \times 500 \times 1$~($ depth \times y-axis \times x-axis$)\\
			2  & 16 Convolutions with size $ 129 \times 1 \times 1$, Batchnorm, ReLu (Spatial filtering layer)\\
			3  & Residual block with N=32 (See Table~\ref{tbl:residualblock}) \\
			4  & Residual block with N=64 (See Table~\ref{tbl:residualblock}) \\
			5  & Fully connected layer with 256 neurons, ReLu \\
			6  & Fully connected layer with 2 neurons (Output layer)\\
		\end{tabular}
	\end{table}
	Table~\ref{tbl:archie} shows our complete model. The input size is the same as the data recordings from ~\cite{kastrati2021eegeyenet}. Along the depth of the input the signals for each EEG electrode is stored and along the y-axis we add the EEG signal per timestamp. Afterwards, our spatial filtering convolution is placed together with a batch normalization and a ReLu. The following to blocks are our residual blocks, which are described in detail in Table~\ref{tbl:residualblock}. The only minor special part of those blocks are the unequally sized convolutions. This designe decision as well as the design decision for the spatial filtering convolution are evaluated in Table~\ref{tbl:design}. The last two layers of our model are fully connected layers with 256 and 2 neurons. The 2 neurons represent the output, which are the x and y coordinates of the gaze. Using two fully connected layers in the end are common for regression problems.
	
	\section{Evaluation}
	In this section we describe the data set we used for our evaluations, the training procedure as well as the parameters, the training of the state-of-the-art models, and our evaluations. We made a comparison to the state-of-the-art in terms of mean absolute error (Table~\ref{tbl:results}) as well as the resource consumption and the runtime (Table~\ref{tbl:params}). In addition, we evaluated our model design decisions in Table~\ref{tbl:design}.
	
	\textbf{Data: } For our evaluations, we used the large data set from EEGEyeNet~\cite{kastrati2021eegeyenet} with 356 subjects. From those 356 subjects are 190 female and 166 male in the age of 18 to 80. Each participant was seated 68 cm from a 24-inch monitor with a resolution of 800~$\times$~600 pixels. For head stabilization, a chin rest was used. For recording, the authors used a 500 Hz, 128-channel EEG Geodesic Hydrocel system with midline central recording reference. The impedance for each electrode was check to be below 40 kOhm. For the eye tracking ground truth, a ET EyeLink 1000 Plus from SR Research video-based eye tracker was used with a sampling rate of 500 Hz. The eye tracker was calibrated with a 9 point calibration, and this calibration was repeated until the error was below 1\textdegree. Since EEG data is contaminated by artifacts produced by environmental factors and  muscular noise like heart signals or sweating. Therefore, the data has to be cleaned or filtered. The authors provided the data with minimal and maximal preprocessing. Minimal preprocessing uses interpolation of bad electrodes, and 40 Hz high-pass as well as 0.5 Hz low-pass filtering. For the maximal preprocessing, a larger portion of the data is interpolated and an independent component analysis (ICA) is applied. We used the minimal processed data as the authors did~\cite{kastrati2021eegeyenet} since it produces a more accurate gaze estimation. The synchronization of the eye tracking data and the EEG data is done with EEG Eye~\cite{dimigen2011coregistration} with a maximal error of 2 ms. In total the authors recorded three tasks, Pro-/Antisaccade, Large Grid, and Visual Symbol Search. The Pro-/Antisaccade experiment was used for estimating left or right direction of the eyes based on the EEG signal. The large Grid was used for the gaze estimation, and we use this data in our paper.
	
	\textbf{Hard/Software setup: } We used a Windows 10 PC with 64 GB DDR4 Ram, an AMD Ryzen 9 3950X 16-Core Processor with 3.50 GHz, and a 1050 TI GPU from NVIDIA. For CUDA we used version 11.6 and for cuDNN the version 8.3.
	
	\textbf{Training parameters: } We used the adam optimizer~\cite{kingma2014adam} with first momentum 0.9, second momentum 0.999, and weight decay $5*10^{-4}$ in combination with the mean squared loss function. The initial learning rate was set to $10^{-4}$ and after 100 epochs the training was stopped. We did not use any data augmentation and a batch size of 64. We used the exact same split for training and testing as the authors of EEGEyeNet did~\cite{kastrati2021eegeyenet}. Additionally, we made evaluations where we randomly created the training and validation set per epoch out of the data from the training and validation split from the authors. This is indicated with a * in Table~\ref{tbl:results}. The advantage of this is that we have more training data and also a validation set to test our generalization. For the state-of-the-art methods, we used the parameters as given by the authors.
	
	\begin{table}
		\centering
		\caption{Evaluation of different desgine decitions of our model. No spatial means, that the first depth wise convolution is removed. Equal convs means, that both convolutions in a residual block have the same size. MAE is the mean absolut error metric in mm. Model parameters are given in millions. The time is measured for 1000 samples processed individually (Batch size one) and given in seconds. All evaluations are with the random training and testing split per epoch.}
		\label{tbl:design}
		\begin{tabular}{c|ccc}
			Method  & MAE & Model parameters ($10^6$) & Time for 1k (sec)\\ \hline
			Ours (no spatial) & 66.67 & 4.24 & 0.891 \\   
			Ours (equal convs) & 54.38 & 2.12 & 0.818 \\   
			Ours (no spatial, equal convs) & 80.42 & 4.28 & 0.919 \\   
			Ours & \textbf{49.99} & \textbf{2.07} & \textbf{0.633} \\ \hline
		\end{tabular}
	\end{table}
	Table~\ref{tbl:design} shows the evaluation of our design decisions. With no spatial, we mean, that the first $1 \times 1 \times depth$ convolution together with the batch normalization and the ReLu is removed (Layer 2 in Table~\ref{tbl:archie}). As can be seen in Table~\ref{tbl:design} this has a high impact on the accuracy of our model (66.67 vs 49.99 MAE). Additionally, the convolution for the spatial filtering also reduces the amount of parameters as well as the runtime. This is because the convolutions, which usually process the output of the spatial filtering convolutions, have now to process more data (All EEG signals along the depth). Therefore, more parameters are necessary as well as more operations to compute the model output. With equal convs we mean that all convolutions in our residual blocks have the same size. This increases the count of parameters as well as the runtime. The impact on the runtime is significantly higher, since the larger convolutions cause more memory conflicts on the GPU (Memory conflict means that the current data has to be fetched from the global memory and is currently not available for the streaming multiprocessor on the GPU). If we remove both the spatial and the unequal convolutions, we get the worst result, the biggest model, and the highest runtime.
	
	\begin{table}
		\centering
		\caption{Comparioson of our model to the state of the art. Each model is evaluated 5 times with the mean absolut error (MAE) metric in mm. For each model we also computed the mean and standard deviation (STD) over the five runs. The results next to Ours use the identical test, validation, and training split as in ~\cite{kastrati2021eegeyenet}. * indicates that we used random validation and training split for each epoch of training.}
		\label{tbl:results}
		\begin{tabular}{c|ccccc|cc}
			Method  & Run 1 & Run 2 & Run 3 & Run 4 & Run 5 & Mean & STD\\ \hline
			CNN & 85.80 & 88.52 & 85.88 & 86.69 & 86.19 & 86.61 & 1.12 \\   
			PyramidalCNN & 89.84 & 90.33 & 89.63 & 90.70 & 90.76 & 90.25 & 0.50 \\   
			EEGNet & 97.14 & 98.46 & 97.58 & 96.23 & 95.77 & 97.03 & 1.07 \\   
			InceptionTime & 89.75 & 89.05 & 88.63 & 88.26 & 86.25 & 88.38 & 1.31 \\   
			Xception & 95.50 & 92.65 & 93.38 & 95.19 & 96.54 & 94.65 & 1.59 \\   
			Ours & 49.99 & 48.92 & 49.05 & 49.63 & 48.45 & 49.20 & 0.60 \\
			Ours* & 47.24 & 47.17 & 47.51 & 47.50 & 47.24 & \textbf{47.33} & \textbf{0.16} \\ \hline
		\end{tabular}
	\end{table}
	Table~\ref{tbl:results} shows the comparison to the state-of-the-art-methods. We used the mean absolute error metric in mm for the comparison of the different models. As can be seen, our model brings a large boost compared to the other approaches in terms of the average MAE as well as for the standard deviation of five runs. The main reason for this is the spatial filtering convolution, as can be seen in Table~\ref{tbl:design}. Additionally, the unequal convolutions together with the spatial filtering convolutions also improve the results significantly (See Table~\ref{tbl:design}). To further improve the results, we also inspected an approach to increase the amount of training data. We used a random train validations split after each epoch, which is indicated by * in Table~\ref{tbl:results}. As can be seen, this improves the MAE slightly, but stabilizes the results between multiple runs. This can be seen by comparing the standard deviation of the five runs between Ours and Ours*. Overall, our model increases the MAE and therefore the EEG-based gaze estimation by a large margin, but it should not be forgotten, that image-based eye tracking is still more accurate. With an error of 1\textdegree, an image-based eye tracker would have a distance error of 11,8 mm and with 2\textdegree, a distance error of 23,7 mm regarding the setup of the recording as described in ~\cite{kastrati2021eegeyenet}. Therefore, EEG-based gaze estimation is still less accurate compared to image-based eye tracking and the data is also recorded under laboratory settings. Nevertheless, we can state that our results are promising and that further research is needed in the field of EEG eye tracking. This is especially true for the usage of dry EEG electrodes since those would be comfortable to the user compared to worn eye trackers, which usually feel like glasses or the adjustable eye cameras limit the field of view.
	
	\begin{table}
		\centering
		\caption{Runtime and model size comparison with the state of the art. The model parameters are given in millions. The time is measured in seconds over the entire test set with a batch size of 64.}
		\label{tbl:params}
		\begin{tabular}{c|cc}
			Method  & Model parameters ($10^6$) & Test set with 64 batch size in seconds\\ \hline
			CNN & 5.10 & 5.34 \\  
			PyramidalCNN & 0.41 & 5.54 \\   
			EEGNet & \textbf{0.03} & 17.60 \\   
			InceptionTime & 2.59 & 7.91 \\   
			Xception & 1.34 & 9.16 \\   
			Ours & 2.07 & \textbf{0.65} \\ \hline
		\end{tabular}
	\end{table}
	Table~\ref{tbl:params} shows a resource comparison to the state-of-the-art methods. For the state-of-the-art methods we used CUDA 11.6 with CUDA enabled, and we also checked that they use the GPU by comparing the runtime of CPU only and CUDA enabled PyTorch~\cite{NEURIPS2019_9015}. The CPU implementation increases the runtime by more than a factor of 2. As can be seen in Table~\ref{tbl:params}, the smallest model is EEGNet with 0.03 million parameters. Ours has 2.07 million parameters but only a runtime of 0.65 seconds on the entire test set with a batch size of 64. Batch size of 64 means, that 64 samples are processed in parallel. The higher throughput of our model compared to the other approaches, steams from multiple sources. First, we do not use large convolutions like the other approaches. The CNN for example uses a convolution kernel of size 64 multiple times. Those large convolutions lead to many memory conflicts on the GPU, which lead to a stop of the processing and reloading of the data from the global GPU memory. This can lead to a nearly serial processing of the data, in which case the CPU would be faster compared to the GPU since the CPU has a higher clock frequency. In addition to the large convolutions, we used dlib~\cite{dlib09} and C++ which is probably much faster compared to python and the precompiled PyTorch implementation. For the runtime evaluation, we measured only the processing calls for our implementation, as well as the implementation of the state-of-the-art. Therefore, the scripting language python itself should not have a large impact on the result, but the authors used many wrapper classes which have an impact. It is also possible, that the official precompiled PyTorch framework needs to reshape the data first before it is transferred to the GPU. The third and last reason for our better throughput, is that our model is short and not deep as the state-of-the-art-method are. Therefore, we have fewer dependencies and more computations can be done in parallel.
	
	\section{Limitations}
	The proposed approach improves the gaze estimation significantly by 3.5 cm, but this results in a total mean absolut error of 4.9 cm which is still a multiple of modern image based eye trackers. The image based eye trackers are usually in the range of 1\textdegree~to 2\textdegree~in studies, while given way better accuracies by the manufacturing company (Usually 0.5\textdegree~or 0.2\textdegree). In addition, the data set which we used for our evaluation is recorded under laboratory conditions, this means in real world studies, our approach would probably be less accurate. Another limitation of our approach is that the data we used was recorded with a conventional EEG, which uses wet electrodes. The procedure to set up a subject with those electrodes is laborious and time-consuming, but the EEG signal has a higher quality compared to dry EEG electrodes~\cite{mathewson2017high}. While the goal is to use the dry EEG, in the end our approach is only evaluated on the normal wet electrodes. 
	
	\section{Conclusion}
	In this paper we proposed a novel DNN architecture for EEG based gaze estimation. Our architecture advances the state of the art by 3.5 cm with a final mean absolut error of 4.9 cm. While this is not as accurate as image based eye trackers (1\textdegree to 2\textdegree in real world studies), it is a huge improvement to the previousely published method~\cite{kastrati2021eegeyenet}. We showed that the spatial convolution has the biggest impact on the model as well as the preceding large convlutions in the residual blocks. For future research it is also important to look into earlier segments of the EEG data, as well as using larger data samples for gaze estimation. We think this will further improve the gaze estimation accuracy and lead to real world EEG based eye tracking.
	
	Other research questions which rise from the possibility of EEG-based eye tracking are the usage of dry EEG electrodes. This is the most important step since the conventional electrodes which require contact material are time-consuming to set up and not comfortable for the subject. Additionally, to the dry electrodes more data is needed, especially out of the lab. There exist many challenges for EEG-based eye tracking like muscle movement, magnetic fields, movement of other humans in the surrounding area, and many more. All of these factors influence the EEG signal, and it is important to know if it is possible to eliminate these effects with modern machine learning approaches like neural networks. While the mentioned tasks would represent a great step towards EEG-based eye tracking, other research topics would arise too. Like a comparison of image-based eye trackers with EEG-based eye tracking in the wild, where the subject would wear both or how the EEG behaves in environments where image-based eye trackers struggle, like strong NIR reflections or nearly closed eyes. Based on those results, a combinatorial solution could also find its way into the world.

	\bibliographystyle{plain}
	\bibliography{template}

\begin{thebibliography}{10}

\bibitem{bellizzi2022innovative}
Lorenzo Bellizzi, Giuseppe Bevilacqua, Valerio Biancalana, Mario Carucci,
  Roberto Cecchi, Piero Chessa, Aniello Donniacuo, Marco Mandal{\`a}, and
  Leonardo Stiaccini.
\newblock An innovative eye-tracker: Main features and demonstrative tests.
\newblock {\em Review of Scientific Instruments}, 93(3):035006, 2022.

\bibitem{blignaut2014mapping}
Pieter Blignaut.
\newblock Mapping the pupil-glint vector to gaze coordinates in a simple
  video-based eye tracker.
\newblock {\em Journal of Eye Movement Research}, 7(1), 2014.

\bibitem{dahmani2020intelligent}
Mahmoud Dahmani, Muhammad~EH Chowdhury, Amith Khandakar, Tawsifur Rahman,
  Khaled Al-Jayyousi, Abdalla Hefny, and Serkan Kiranyaz.
\newblock An intelligent and low-cost eye-tracking system for motorized
  wheelchair control.
\newblock {\em Sensors}, 20(14):3936, 2020.

\bibitem{dimigen2011coregistration}
Olaf Dimigen, Werner Sommer, Annette Hohlfeld, Arthur~M Jacobs, and Reinhold
  Kliegl.
\newblock Coregistration of eye movements and eeg in natural reading: analyses
  and review.
\newblock {\em Journal of experimental psychology: General}, 140(4):552, 2011.

\bibitem{ding2019classifying}
Xinfang Ding, Xinxin Yue, Rui Zheng, Cheng Bi, Dai Li, and Guizhong Yao.
\newblock Classifying major depression patients and healthy controls using eeg,
  eye tracking and galvanic skin response data.
\newblock {\em Journal of affective Disorders}, 251:156--161, 2019.

\bibitem{dosovitskiy2020image}
Alexey Dosovitskiy, Lucas Beyer, Alexander Kolesnikov, Dirk Weissenborn,
  Xiaohua Zhai, Thomas Unterthiner, Mostafa Dehghani, Matthias Minderer, Georg
  Heigold, Sylvain Gelly, et~al.
\newblock An image is worth 16x16 words: Transformers for image recognition at
  scale.
\newblock {\em arXiv preprint arXiv:2010.11929}, 2020.

\bibitem{duchowski2017eye}
Andrew~T Duchowski and Andrew~T Duchowski.
\newblock {\em Eye tracking methodology: Theory and practice}.
\newblock Springer, 2017.

\bibitem{eivazi2018eyemic}
Shaharam Eivazi and Maximilian Maurer.
\newblock Eyemic: an eye tracker for surgical microscope.
\newblock In {\em Proceedings of the 2018 ACM Symposium on Eye Tracking
  Research \& Applications}, pages 1--2, 2018.

\bibitem{ETRA2018FuhlW}
W.~Fuhl, S.~Eivazi, B.~Hosp, A.~Eivazi, W.~Rosenstiel, and E.~Kasneci.
\newblock Bore: Boosted-oriented edge optimization for robust, real time remote
  pupil center detection.
\newblock In {\em Eye Tracking Research and Applications, ETRA}, 2018.

\bibitem{ICCVW2019FuhlW}
W.~Fuhl, D.~Geisler, W.~Rosenstiel, and E.~Kasneci.
\newblock The applicability of cycle gans for pupil and eyelid segmentation,
  data generation and image refinement.
\newblock In {\em International Conference on Computer Vision Workshops,
  ICCVW}, 11 2019.

\bibitem{CAIP2019FuhlW}
W.~Fuhl, W.~Rosenstiel, and E.~Kasneci.
\newblock 500,000 images closer to eyelid and pupil segmentation.
\newblock In {\em Computer Analysis of Images and Patterns, CAIP}, 11 2019.

\bibitem{WTDTE022017}
W.~Fuhl, T.~Santini, D.~Geisler, T.~C. Kübler, and E.~Kasneci.
\newblock Eyelad: Remote eye tracking image labeling tool.
\newblock In {\em 12th Joint Conference on Computer Vision, Imaging and
  Computer Graphics Theory and Applications (VISIGRAPP 2017)}, 02 2017.

\bibitem{WTDTWE092016}
W.~Fuhl, T.~Santini, D.~Geisler, T.~C. Kübler, W.~Rosenstiel, and E.~Kasneci.
\newblock Eyes wide open? eyelid location and eye aperture estimation for
  pervasive eye tracking in real-world scenarios.
\newblock In {\em ACM International Joint Conference on Pervasive and
  Ubiquitous Computing: Adjunct publication -- PETMEI 2016}, 09 2016.

\bibitem{WTE032017}
W.~Fuhl, T.~Santini, and E.~Kasneci.
\newblock Fast and robust eyelid outline and aperture detection in real-world
  scenarios.
\newblock In {\em IEEE Winter Conference on Applications of Computer Vision
  (WACV 2017)}, 03 2017.

\bibitem{UMUAI2020FUHL}
Wolfgang Fuhl.
\newblock From perception to action using observed actions to learn gestures.
\newblock {\em User Modeling and User-Adapted Interaction}, pages 1--18, 08
  2020.

\bibitem{fuhl2022groupgazer}
Wolfgang Fuhl.
\newblock Groupgazer: A tool to compute the gaze per participant in groups with
  integrated calibration to map the gaze online to a screen or beamer
  projection.
\newblock {\em arXiv preprint arXiv:2201.07692}, 2022.

\bibitem{fuhl2015excuse}
Wolfgang Fuhl, Thomas K{\"u}bler, Katrin Sippel, Wolfgang Rosenstiel, and
  Enkelejda Kasneci.
\newblock Excuse: Robust pupil detection in real-world scenarios.
\newblock In {\em International conference on computer analysis of images and
  patterns}, pages 39--51. Springer, 2015.

\bibitem{fuhl2017fast}
Wolfgang Fuhl, Thiago Santini, and Enkelejda Kasneci.
\newblock Fast camera focus estimation for gaze-based focus control.
\newblock {\em arXiv preprint arXiv:1711.03306}, 2017.

\bibitem{fuhl2022pistol}
Wolfgang Fuhl, Daniel Weber, and Enkelejda Kasneci.
\newblock Pistol: Pupil invisible supportive tool to extract pupil, iris, eye
  opening, eye movements, pupil and iris gaze vector, and 2d as well as 3d
  gaze.
\newblock {\em arXiv preprint arXiv:2201.06799}, 2022.

\bibitem{gibaldi2017evaluation}
Agostino Gibaldi, Mauricio Vanegas, Peter~J Bex, and Guido Maiello.
\newblock Evaluation of the tobii eyex eye tracking controller and matlab
  toolkit for research.
\newblock {\em Behavior research methods}, 49(3):923--946, 2017.

\bibitem{hosp2020remoteeye}
Benedikt Hosp, Shahram Eivazi, Maximilian Maurer, Wolfgang Fuhl, David Geisler,
  and Enkelejda Kasneci.
\newblock Remoteeye: An open-source high-speed remote eye tracker.
\newblock {\em Behavior research methods}, 52(3):1387--1401, 2020.

\bibitem{hua2006video}
Hong Hua, Prasanna Krishnaswamy, and Jannick~P Rolland.
\newblock Video-based eyetracking methods and algorithms in head-mounted
  displays.
\newblock {\em Optics Express}, 14(10):4328--4350, 2006.

\bibitem{kamangar2020literature}
Arash Kamangar.
\newblock A literature review of customer behaviour patterns on e-commerce
  websites using an eye tracker.
\newblock {\em The Marketing Review}, 20(1-2):73--91, 2020.

\bibitem{kassner2014pupil}
Moritz Kassner, William Patera, and Andreas Bulling.
\newblock Pupil: an open source platform for pervasive eye tracking and mobile
  gaze-based interaction.
\newblock In {\em Proceedings of the 2014 ACM international joint conference on
  pervasive and ubiquitous computing: Adjunct publication}, pages 1151--1160,
  2014.

\bibitem{kastrati2021using}
Ard Kastrati, Martyna~Beata Plomecka, Roger Wattenhofer, and Nicolas Langer.
\newblock Using deep learning to classify saccade direction from brain
  activity.
\newblock In {\em ACM Symposium on Eye Tracking Research and Applications},
  pages 1--6, 2021.

\bibitem{kastrati2021eegeyenet}
Ard Kastrati, Martyna Martyna~Beata P{\l}omecka, Dami{\'a}n Pascual, Lukas
  Wolf, Victor Gillioz, Roger Wattenhofer, and Nicolas Langer.
\newblock Eegeyenet: a simultaneous electroencephalography and eye-tracking
  dataset and benchmark for eye movement prediction.
\newblock {\em arXiv preprint arXiv:2111.05100}, 2021.

\bibitem{dlib09}
Davis~E. King.
\newblock Dlib-ml: A machine learning toolkit.
\newblock {\em Journal of Machine Learning Research}, 10:1755--1758, 2009.

\bibitem{kingma2014adam}
Diederik~P Kingma and Jimmy Ba.
\newblock Adam: A method for stochastic optimization.
\newblock {\em arXiv preprint arXiv:1412.6980}, 2014.

\bibitem{liu2022convnet}
Zhuang Liu, Hanzi Mao, Chao-Yuan Wu, Christoph Feichtenhofer, Trevor Darrell,
  and Saining Xie.
\newblock A convnet for the 2020s.
\newblock In {\em Proceedings of the IEEE/CVF Conference on Computer Vision and
  Pattern Recognition}, pages 11976--11986, 2022.

\bibitem{lopez2014dry}
Miguel~Angel Lopez-Gordo, Daniel Sanchez-Morillo, and F~Pelayo Valle.
\newblock Dry eeg electrodes.
\newblock {\em Sensors}, 14(7):12847--12870, 2014.

\bibitem{mantiuk2012yourself}
Rados{\l}aw Mantiuk, Micha{\l} Kowalik, Adam Nowosielski, and Bartosz Bazyluk.
\newblock Do-it-yourself eye tracker: Low-cost pupil-based eye tracker for
  computer graphics applications.
\newblock In {\em International Conference on Multimedia Modeling}, pages
  115--125. Springer, 2012.

\bibitem{massin2020development}
Lo{\"\i}c Massin, Vincent Nourrit, Cyril Lahuec, Fabrice Seguin, Laure Adam,
  Emmanuel Daniel, et~al.
\newblock Development of a new scleral contact lens with encapsulated
  photodetectors for eye tracking.
\newblock {\em Optics Express}, 28(19):28635--28647, 2020.

\bibitem{mathewson2017high}
Kyle~E Mathewson, Tyler~JL Harrison, and Sayeed~AD Kizuk.
\newblock High and dry? comparing active dry eeg electrodes to active and
  passive wet electrodes.
\newblock {\em Psychophysiology}, 54(1):74--82, 2017.

\bibitem{munoz2019measuring}
Francisco Mu{\~n}oz-Leiva, Janet Hern{\'a}ndez-M{\'e}ndez, and Diego
  G{\'o}mez-Carmona.
\newblock Measuring advertising effectiveness in travel 2.0 websites through
  eye-tracking technology.
\newblock {\em Physiology \& behavior}, 200:83--95, 2019.

\bibitem{nagel2019world}
Sebastian Nagel and Martin Sp{\"u}ler.
\newblock World’s fastest brain-computer interface: combining eeg2code with
  deep learning.
\newblock {\em PloS one}, 14(9):e0221909, 2019.

\bibitem{NEURIPS2019_9015}
Adam Paszke, Sam Gross, Francisco Massa, Adam Lerer, James Bradbury, Gregory
  Chanan, Trevor Killeen, Zeming Lin, Natalia Gimelshein, Luca Antiga, Alban
  Desmaison, Andreas Kopf, Edward Yang, Zachary DeVito, Martin Raison, Alykhan
  Tejani, Sasank Chilamkurthy, Benoit Steiner, Lu~Fang, Junjie Bai, and Soumith
  Chintala.
\newblock Pytorch: An imperative style, high-performance deep learning library.
\newblock In {\em Advances in Neural Information Processing Systems 32}, pages
  8024--8035. Curran Associates, Inc., 2019.

\bibitem{ramanauskas2006calibration}
Nerijus Ramanauskas.
\newblock Calibration of video-oculographical eye-tracking system.
\newblock {\em Elektronika Ir Elektrotechnika}, 72(8):65--68, 2006.

\bibitem{santini2017eyerectoo}
Thiago Santini, Wolfgang Fuhl, David Geisler, and Enkelejda Kasneci.
\newblock Eyerectoo: Open-source software for real-time pervasive head-mounted
  eye tracking.
\newblock In {\em VISIGRAPP (6: VISAPP)}, pages 96--101, 2017.

\bibitem{shishkin2016eeg}
Sergei~L Shishkin, Yuri~O Nuzhdin, Evgeny~P Svirin, Alexander~G Trofimov,
  Anastasia~A Fedorova, Bogdan~L Kozyrskiy, and Boris~M Velichkovsky.
\newblock Eeg negativity in fixations used for gaze-based control: Toward
  converting intentions into actions with an eye-brain-computer interface.
\newblock {\em Frontiers in neuroscience}, 10:528, 2016.

\bibitem{sundstedt2012gazing}
Veronica Sundstedt.
\newblock Gazing at games: An introduction to eye tracking control.
\newblock {\em Synthesis Lectures on Computer Graphics and Animation},
  5(1):1--113, 2012.

\bibitem{swirski2012robust}
Lech {\'S}wirski, Andreas Bulling, and Neil Dodgson.
\newblock Robust real-time pupil tracking in highly off-axis images.
\newblock In {\em Proceedings of the symposium on eye tracking research and
  applications}, pages 173--176, 2012.

\bibitem{swirski2013fully}
Lech Swirski and Neil Dodgson.
\newblock A fully-automatic, temporal approach to single camera, glint-free 3d
  eye model fitting.
\newblock {\em Proc. PETMEI}, pages 1--11, 2013.

\bibitem{tonsen2020high}
Marc Tonsen, Chris~Kay Baumann, and Kai Dierkes.
\newblock A high-level description and performance evaluation of pupil
  invisible.
\newblock {\em arXiv preprint arXiv:2009.00508}, 2020.

\bibitem{valtakari2021eye}
Niilo~V Valtakari, Ignace~TC Hooge, Charlotte Viktorsson, P{\"a}r Nystr{\"o}m,
  Terje Falck-Ytter, and Roy~S Hessels.
\newblock Eye tracking in human interaction: Possibilities and limitations.
\newblock {\em Behavior Research Methods}, 53(4):1592--1608, 2021.

\bibitem{vettori2020combined}
Sofie Vettori, Stephanie Van~der Donck, Jannes Nys, Pieter Moors, Tim
  Van~Wesemael, Jean Steyaert, Bruno Rossion, Milena Dzhelyova, and Bart Boets.
\newblock Combined frequency-tagging eeg and eye-tracking measures provide no
  support for the “excess mouth/diminished eye attention” hypothesis in
  autism.
\newblock {\em Molecular autism}, 11(1):1--22, 2020.

\bibitem{vetturi2020use}
David Vetturi, Michela Tiboni, Giulio Maternini, and Michela Bonera.
\newblock Use of eye tracking device to evaluate the driver’s behaviour and
  the infrastructures quality in relation to road safety.
\newblock {\em Transportation research procedia}, 45:587--595, 2020.

\bibitem{villanueva2004eye}
Arantxa Villanueva, Rafael Cabeza, and Sonia Porta.
\newblock Eye tracking system model with easy calibration.
\newblock In {\em Proceedings of the 2004 symposium on Eye tracking research \&
  applications}, pages 55--55, 2004.

\bibitem{walton2021beyond}
David~R Walton, Rafael~Kuffner Dos~Anjos, Sebastian Friston, David Swapp, Kaan
  Ak{\c{s}}it, Anthony Steed, and Tobias Ritschel.
\newblock Beyond blur: Real-time ventral metamers for foveated rendering.
\newblock {\em ACM Transactions on Graphics}, 40(4):1--14, 2021.

\bibitem{wanluk2016smart}
Nutthanan Wanluk, Sarinporn Visitsattapongse, Aniwat Juhong, and C~Pintavirooj.
\newblock Smart wheelchair based on eye tracking.
\newblock In {\em 2016 9th Biomedical Engineering International Conference
  (BMEiCON)}, pages 1--4. IEEE, 2016.

\bibitem{whitmire2016eyecontact}
Eric Whitmire, Laura Trutoiu, Robert Cavin, David Perek, Brian Scally, James
  Phillips, and Shwetak Patel.
\newblock Eyecontact: scleral coil eye tracking for virtual reality.
\newblock In {\em Proceedings of the 2016 ACM International Symposium on
  Wearable Computers}, pages 184--191, 2016.

\bibitem{wood2016learning}
Erroll Wood, Tadas Baltru{\v{s}}aitis, Louis-Philippe Morency, Peter Robinson,
  and Andreas Bulling.
\newblock Learning an appearance-based gaze estimator from one million
  synthesised images.
\newblock In {\em Proceedings of the Ninth Biennial ACM Symposium on Eye
  Tracking Research \& Applications}, pages 131--138, 2016.

\bibitem{wu2017spatial}
Dongrui Wu, Jung-Tai King, Chun-Hsiang Chuang, Chin-Teng Lin, and Tzyy-Ping
  Jung.
\newblock Spatial filtering for eeg-based regression problems in
  brain--computer interface (bci).
\newblock {\em IEEE Transactions on Fuzzy Systems}, 26(2):771--781, 2017.

\bibitem{xu2018gaze}
Yanyu Xu, Yanbing Dong, Junru Wu, Zhengzhong Sun, Zhiru Shi, Jingyi Yu, and
  Shenghua Gao.
\newblock Gaze prediction in dynamic 360 immersive videos.
\newblock In {\em proceedings of the IEEE Conference on Computer Vision and
  Pattern Recognition}, pages 5333--5342, 2018.

\bibitem{zhang2015appearance}
Xucong Zhang, Yusuke Sugano, Mario Fritz, and Andreas Bulling.
\newblock Appearance-based gaze estimation in the wild.
\newblock In {\em Proceedings of the IEEE conference on computer vision and
  pattern recognition}, pages 4511--4520, 2015.

\end{thebibliography}

\end{document}